\documentclass[fleqn,twoside]{article}
\usepackage{espcrc2}

\usepackage{graphicx}

\newcommand{\bel}{\begin{equation}}
\newcommand{\ee}{\end{equation}}
\def\rys#1#2{\begin{figure}[h]
      \vskip 3mm
      \centerline{
      \includegraphics*[width=0.5 \textwidth]{#1}
      }
      \caption{#2}
      \vskip 3mm
      \end{figure}
      }

\newcommand{\AmS}{{\protect\the\textfont2
  A\kern-.1667em\lower.5ex\hbox{M}\kern-.125emS}}

\font\wielkie=cmr10 scaled\magstep3

\font\duze=cmr10 scaled\magstep1

\begin{document}
\title{}\author{}

\centerline {\wielkie On two weak CC $\Delta$ production models}

\vskip 3truecm

\centerline{\duze Jan T. Sobczyk}

\bigskip

\centerline {\duze Institute of Theoretical Physics}

\medskip

\centerline{ \duze Wroc\l aw University.}
\medskip
\centerline{ \duze pl. M. Borna 9, 50-204 Wroc\l aw, Poland}

\vskip 3truecm

{\bf ABSTRACT}

\smallskip

We perform a detail analysis of two models of neutrino CC $\Delta$
production on free nucleons. First model is a standard one based
on nucleon-Delta transition current with several form-factors.
Second model is a starting point for a construction of Marteau
model with sophisticated analytical computations of nuclear
effects. We conclude that both models lead to similar results.


\bigskip
SHORT TITLE: On two weak CC $\Delta$ production models

\maketitle

\section{INTRODUCTION}
Future precise neutrino measurements e.g. of $\theta_{13}$ require
better understanding of neutrino interactions in few GeV region
\cite{nuint}. In general, in description of neutrino-nucleon
interaction vertex three processes are distinguished:
quasielastic, single pion production (resonance excitation region)
and more inelastic processes taken into account by the formalism
of deep inelastic scattering. Weak single pion production is
therefore a part of more complicated dynamics. It gives an
important contribution to cross section in 1 GeV region and has to
be treated with care. In the past it was a subject of many
theoretical studies \cite{stare}. A sample of existing
experimental data is not conclusive as measurements were made with
a typical precision of 20-25\% \cite{exp}. From a point of view of
Monte Carlo simulation codes there seems to be an agreement that
Rein-Sehgal \cite{reinsehgal} model is most reliable. It includes
contributions from 18 resonances with masses up to $2\ GeV$, their
interference terms together with a non-resonance background.
Recent developments in quark-hadron duality suggest however that
there is no need to consider so many resonances: contributions
from most of them can be described in average by suitably modified
PDF's \cite{bodek}. When reaction takes place on nuclear targets
resonance contributions are additionally smeared out by Fermi
motion. A conclusion is that probably only the $\Delta$ excitation
has to be treated independently \cite{casper}.

One way to describe $\Delta$ excitation is to construct a current
$<\Delta |J^{\mu}|N>$ with phenomenological form-factor
constrained by CVC and PCAC arguments \cite{ff}. There have been
also attempts to calculate such form-factors from first principles
in the quark model \cite{quark}. A precision with which
form-factors are known cannot be better then experimental
uncertainties.

Few authors tried to discuss nuclear effects in single pion
production in a framework of more systematic theoretical schemes.
One of such models was developed by Marteau \cite{marteau}. It
includes: Fermi motion, Pauli blocking, elementary 1p-1h,
1$\Delta$-1h and 2p-2h excitations, modification of $\Delta$ width
in a nuclear matter, RPA corrections and finite volume effects.
Before all the nuclear effects are taken into account, the model
is based on simplified dynamical assumptions about CC neutrino
$\Delta$ excitation on {\it free} nucleons. These simplifications
are necessary in order to perform calculations of nuclear effects
in compact and elegant way.

\rys{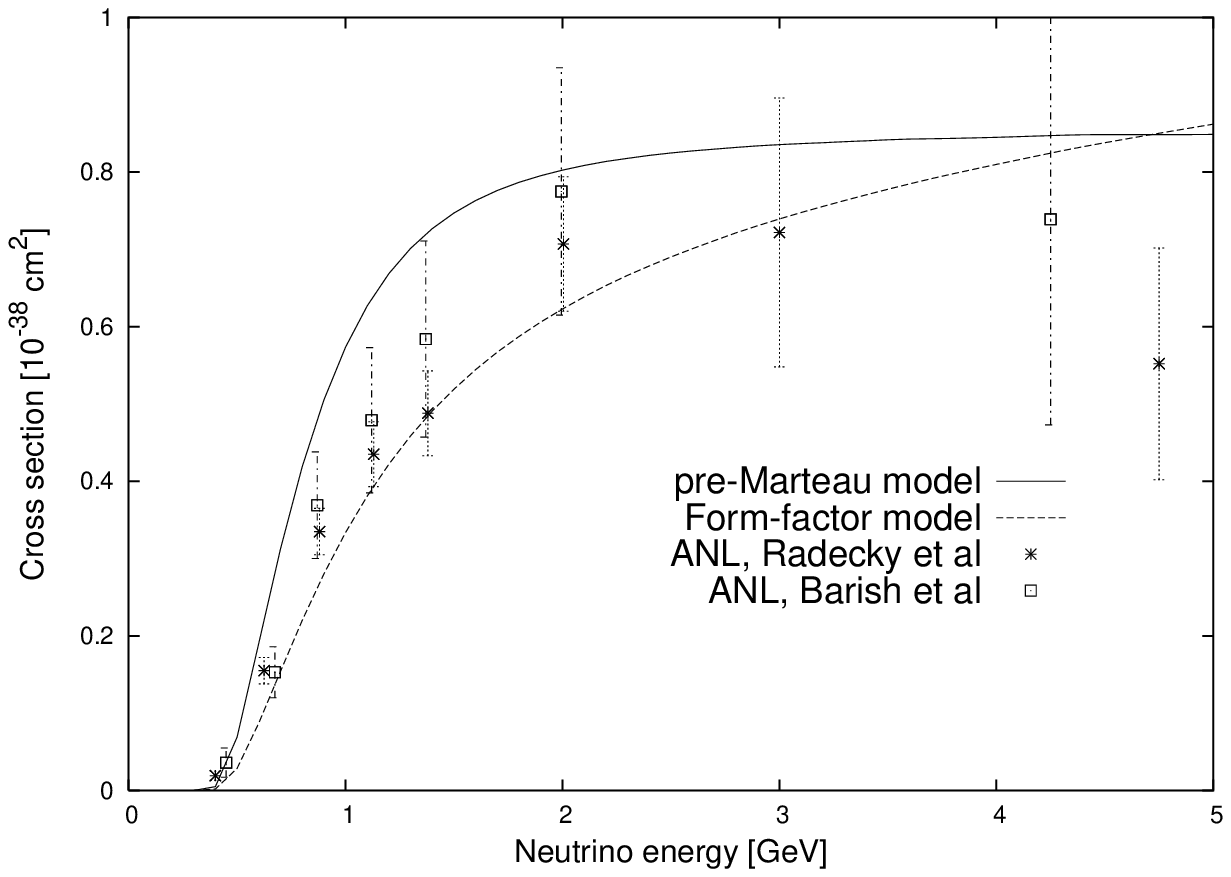} {Total cross section for $\pi^+$ production
via $\Delta^{++}$ excitation. Experimental points are taken from
\cite{exp}. No constraints on hadronic invariant mass are
imposed.}

\rys{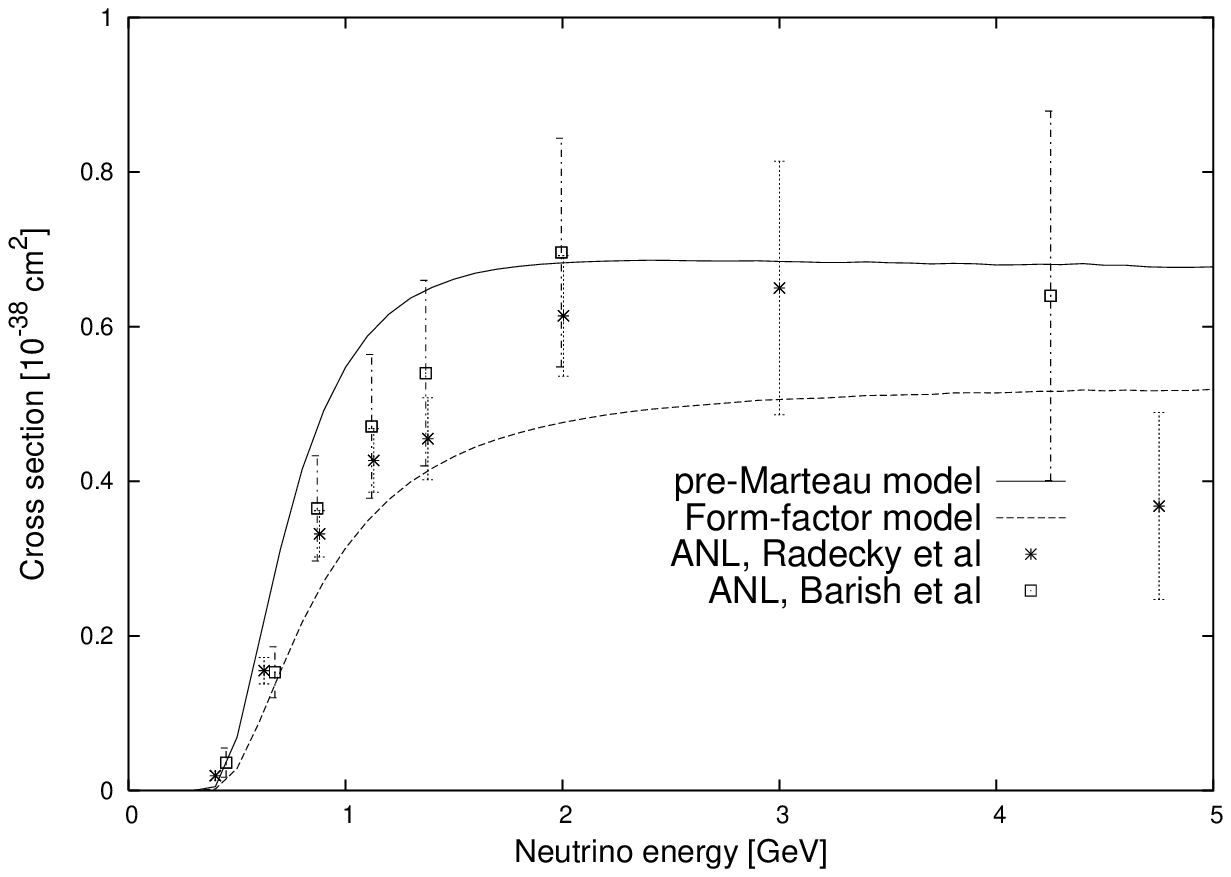} {Total cross section of $\pi^+$ production
via $\Delta^{++}$ excitation with a restriction on invariant
hadronic mass $W\leq 1.4 GeV$. Experimental points are taken from
\cite{exp} .}

\rys{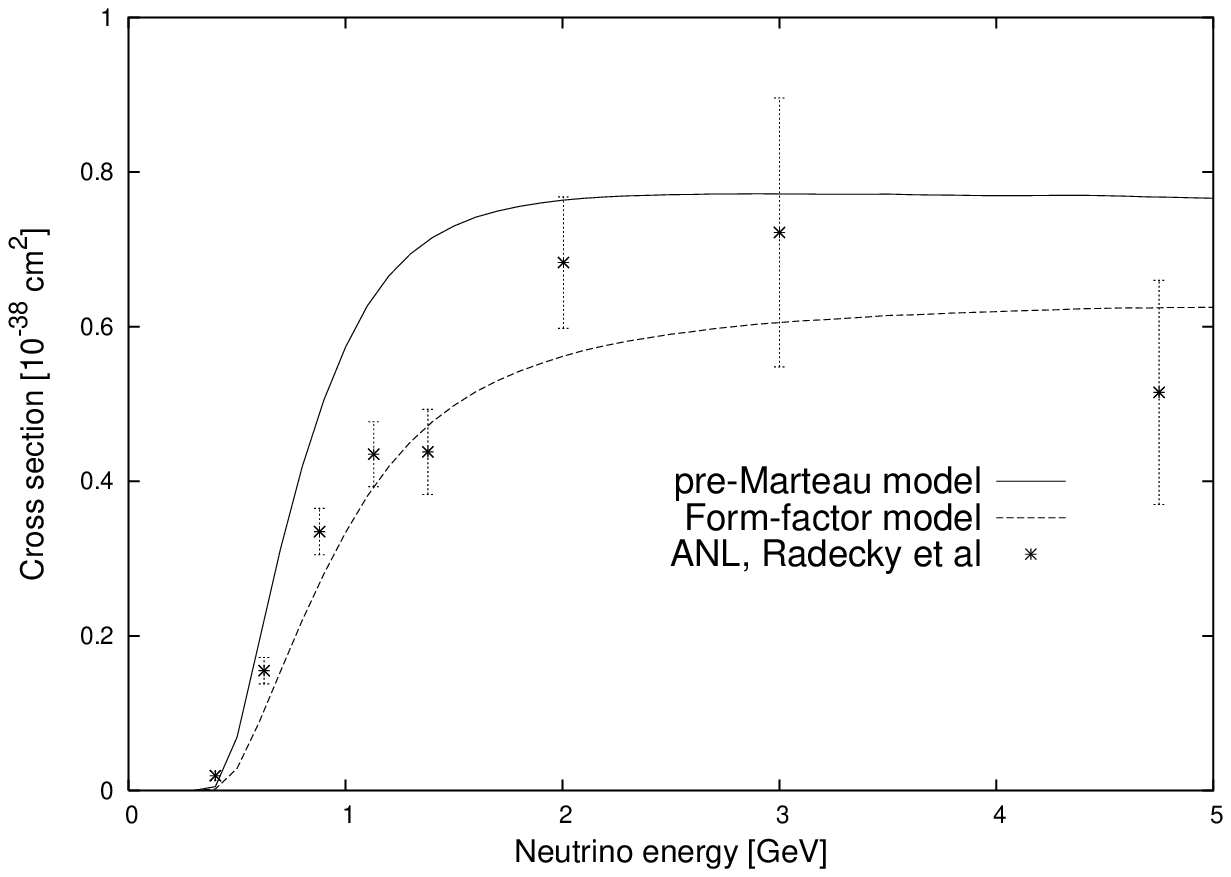} {Total cross section of $\pi^+$ production
via $\Delta^{++}$ excitation with a restriction on invariant
hadronic mass $W\leq 1.6 GeV$. Experimental points are taken from
\cite{exp} .}

\rys{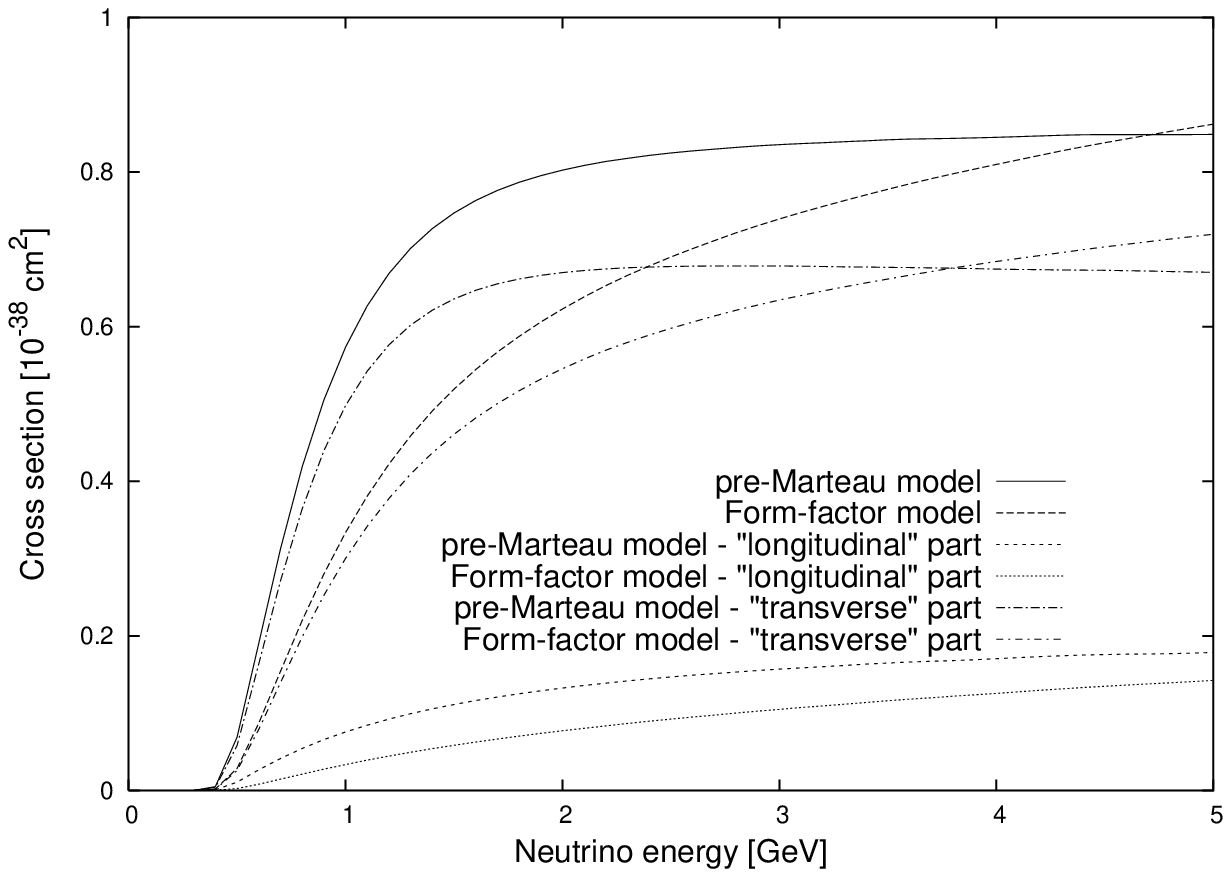} {Total cross section of $\pi^+$ production
via $\Delta^{++}$ excitation with a decomposition into
longitudinal and transverse contributions.}

\rys{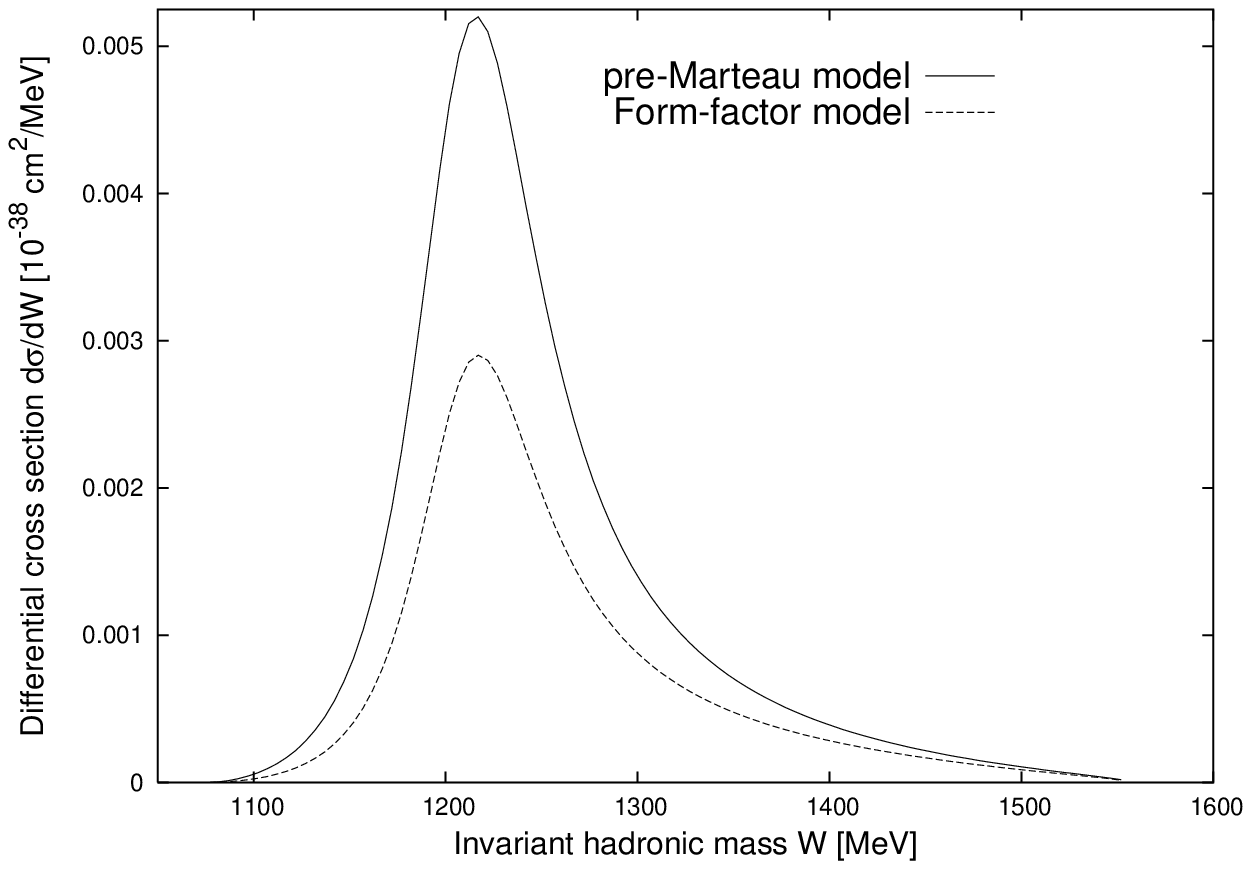} {Differential cross section of
invariant hadronic mass for $\pi^+$ production via $\Delta^{++}$
excitation. Neutrino energy is $E_{\nu}=1 GeV$. }

\rys{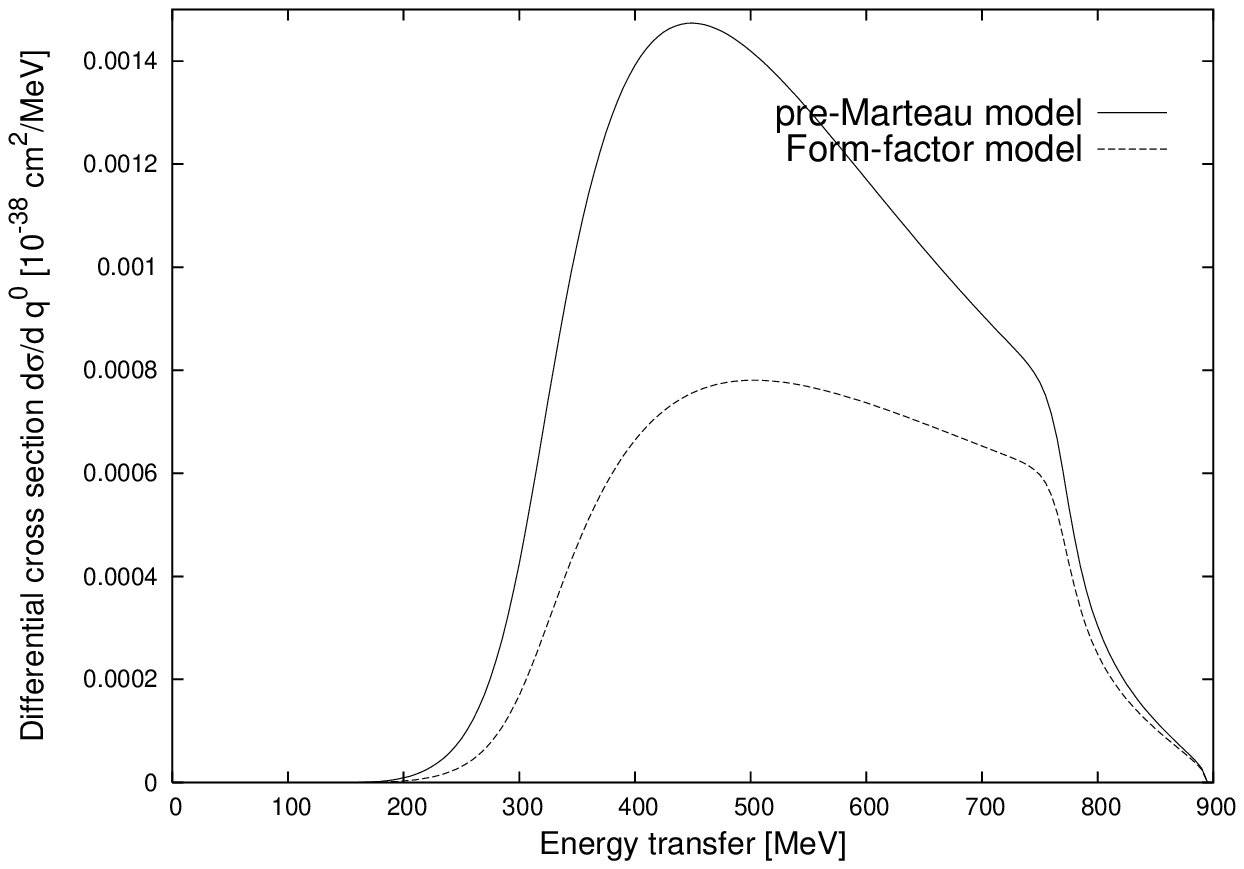}
{Differential cross section of energy transfer for $\pi^+$
production via $\Delta^{++}$ excitation. Neutrino energy is
$E_{\nu}=1 GeV$.}

\rys{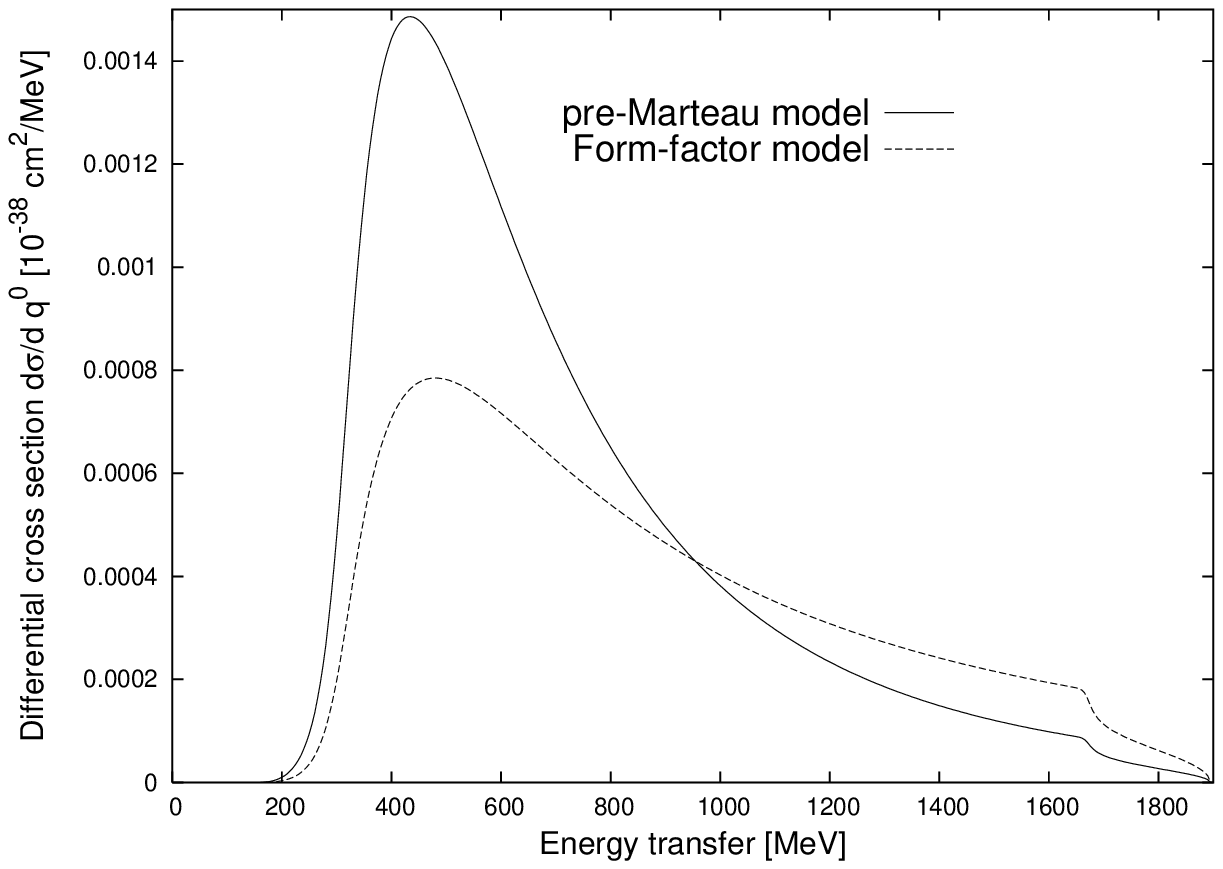} {Differential cross section of energy
transfer for $\pi^+$ production via $\Delta^{++}$ excitation.
Neutrino energy is $E_{\nu}=2 GeV$.}

\rys{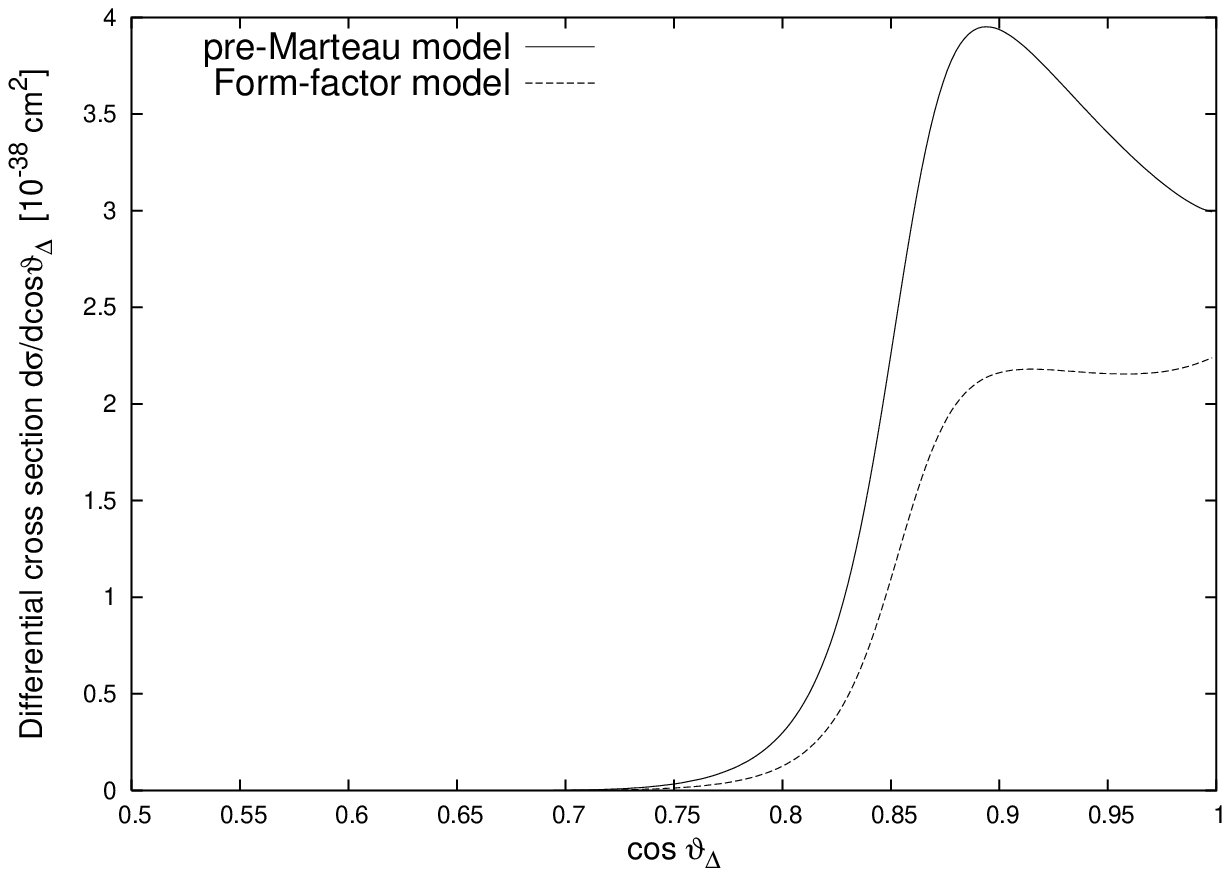} {Differential cross section of $\cos
\Theta_{\Delta}$  for $\pi^+$ production via $\Delta^{++}$
excitation. Neutrino energy is $E_{\nu}=1 GeV$.}

\rys{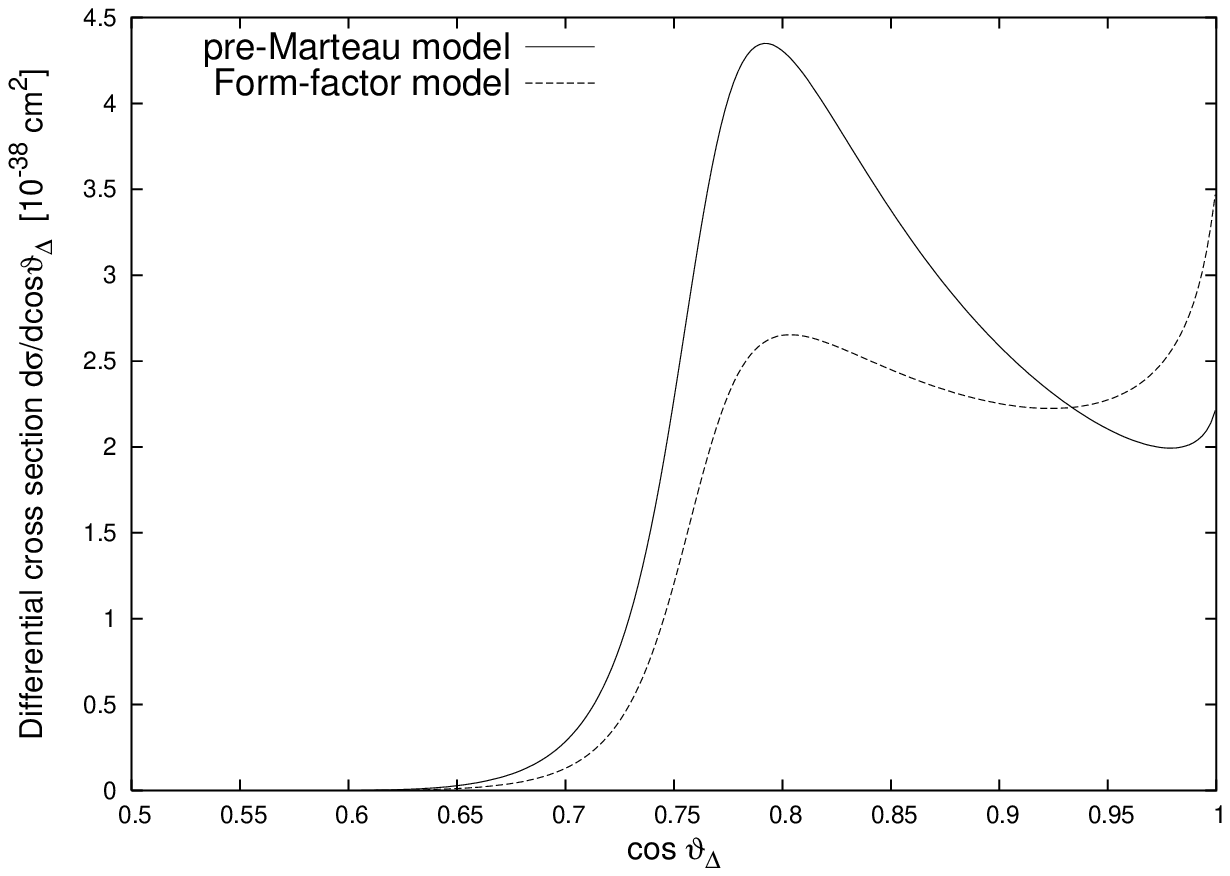} {Differential cross section of $\cos
\Theta_{\Delta}$  for $\pi^+$ production via $\Delta^{++}$
excitation. Neutrino energy is $E_{\nu}=2 GeV$.}

The Marteau model has been discussed in both experimental and
theoretical  context. On experimental side \cite{mar_exp} the
model is used in the Monte Carlo code of K2K collaboration to give
predictions for NC $\pi^0$ production. There is an open dispute as
its predictions are much smaller then those of Rein-Seghal model
for coherent pion production \cite{RS_NC}. It has been also
investigated for the problem of pion-less $\Delta$ decays in the
nuclear matter \cite{casper}. The Marteau model was mentioned by
several authors dealing with theoretical description of nuclear
effects in neutrino interactions in $GeV$ region \cite{mar_theo}.

Due to the lack of experimental data it is difficult to evaluate
how successful the model is in describing nuclear effects.
Certainly it is important to investigate its properties by
performing comparisons with other approaches. An important test is
to investigate basic dynamical assumptions of the model and
compare results with predictions of other models and with existing
experimental data on free nucleons. Such tests are presented in
this paper. Our main conclusion is that CC neutrino-{\it free}
nucleon interaction model being a basis for the Marteau model (we
call that model {\it pre-Marteau model}) leads to very similar
results as the model \cite{ff}.

\section{RESULTS}

Marteau model \cite{marteau} is a sophisticated model containing in one
theoretical scheme neutrino CC quasielastic and $\Delta$ excitation reactions
on nuclei. We present below its basic
assumption for the
differential cross section for CC neutrino
$\Delta^{++}$ excitation on {\it free}
nucleon:

\bel\begin{array}{l}\displaystyle d^2\sigma_M^{\Delta^{++}}=
{G^2\cos^2\theta_c q M_{\Delta}^3 \Gamma_{\Delta}\over
24\pi^2 E^2(M+\omega ) }\\[4mm]
\displaystyle
\qquad\times {L_{\mu\nu}(\tilde
H^{\mu\nu})_{\vec p=0} d\omega
dq\over
\bigl((M+\omega )^2-\vec
q^2-M_{\Delta}^2\bigr)^2+M_{\Delta}^2\Gamma_{\Delta}^2} \end{array}\ee

In order to avoid a confusion we call this model {\it pre-Marteau
model}. The above formula is derived in Appendix A, also the
notation is explained there in detail.

In this paper we compare predictions of pre-Marteau model with the model for
neutrino CC $\Delta$ excitation based on nucleon-$\Delta$ transition current
with several form-factors. In the latter model \cite{ff} (we call it
Form-Factor Model) one calculates cross section
in a standard way. Straightforward computations lead to expression:

\bel\begin{array}{l}\displaystyle d^2\sigma_{FF}^{\Delta^{++}}=
{G^2\cos^2\theta_c\Gamma_{\Delta}\over 64\pi^2E^2}{L_{\mu\nu}{\cal
H}^{\mu\nu}q d\omega dq\over (W-M_{\Delta}^2)
+\Gamma_{\Delta}^2/4}\end{array}\ee

In order to calculate ${\cal H}^{\mu\nu}$ one introduces
nucleon-$\Delta$ transition current \cite{ls}

$${\cal J}^{\alpha}=\sqrt{3}\ \overline{\Psi_{\mu}}(p')\Bigl( \bigl({C^V_3(Q^2)\over
M}(g^{\mu\alpha}\hat q-q^{\mu}\gamma^{\alpha})$$
$$ + {C^V_4(Q^2)\over
M^2}(g^{\mu\alpha}q\cdot p'-q^{\mu}p'^{\alpha})\bigr)\gamma_5 +$$
\bel\begin{array}{l}\displaystyle
\qquad+{C^A_4(Q^2)\over M^2}(g^{\mu\alpha}q\cdot
p'-q^{\mu}p'^{\alpha})\\[4mm]
\displaystyle
+C^A_5(Q^2)g^{\mu\alpha}+{C^A_6(Q^2)\over
M^2}q^{\mu}q^{\alpha}\Bigr)u(p). \end{array}\ee

where $q^{\mu}=p^{\mu}-p'^{\mu}$. A factor $\sqrt{3}$ is present
in the current for $\Delta^{++}$ production. In our numerical
calculations we used the form factors from \cite{singh}. Equations
(1) and (2) as they are written very similar. In fact the
kinematics is the same in both cases. The inequivalence of two
expressions comes from hadronic tensors $(\tilde H^{\mu\nu})_{\vec
p=0}$ ${\cal H}^{\mu\nu}$ which are calculated in different ways.

A comparison of predictions of two models was done for a {\it free} nucleon
target
because the existing experimental data applies to this situation.
We compared total cross section in the energy range up to $5\ GeV$
with data from \cite{exp}. We presented plots without
bound on the invariant hadronic mass and with bounds $1.4\ GeV$ and $1.6\
GeV$ as such experimental data is available.

We made a decomposition of the total cross section into two parts:
"transverse" and "longitudinal" according to spin-isospin operators present
in the hadronic current. In the pre-Marteau model longitudinal operators are
present only in $H_{00}, H_{33}, H_{03}$ and transverse only in
$H_{11}, H_{12}$ (in the frame $\vec q=(0,0,q)$).
Therefore we define "longitudinal" and "transverse" parts as those coming
from
corresponding terms in $L_{\mu\nu}H^{\mu\nu}$.

For two theoretical models we performed also a comparison of
differential cross sections for neutrino energy of $1 GeV$ and $2
GeV$ of: invariant mass, energy transfer and
$\cos\theta_{\Delta}$, an angle between momenta of incident
neutrino and $\Delta$.

\medskip
Our conclusions are the following:

i) For neutrino energy lower than $E_{\nu}=5 GeV$ (Fig. 1)
pre-Marteau model gives rise to higher values of the total cross
section but both models approximately agree with the existing
experimental data.

ii) For $E_{\nu}=5 GeV$ Form-Factor Model predicts the values of
the total cross section greater than pre-Marteau model.
Pre-Marteau model cross section becomes approximately constant
while Form-Factor Model cross section still increases with
neutrino energy.

iii) With constraints on the value of invariant hadronic mass
(Figs 2, 3) both models give rise to predictions similar in shape and in
agreement with
experimental data.
Values of total cross section of
pre-Marteau model are about 20\% higher than predicted by Form-factor model.

iv) Contributions from "longitudinal" and "transverse"
contributions (Fig. 4) to the total cross section are similar in
both models.

v) Invariant hadronic mass distributions (Fig. 5, 6) are very
similar, the only difference is in scale and comes from different
values of the total cross section. For higher values of neutrino
energy Form-Factor Model gives rise to greater contribution from
higher values of $W$. This is why cuts on the invariant mass are
more restrictive for that model.

vi) Differential cross sections of energy transfer (Fig. 7, 8) for
both models are similar in shape. There is an surprising decline
of differential cross section in energy transfer at about $\nu
=0.7 GeV$ for $E_{\nu}=1 GeV$ (Fig. 7) and at about $\nu =1.7 GeV$
for $E_{\nu}=2 GeV$ (Fig. 8) present in predictions of both
models. A possible explanation is kinematical in origin. For a
fixed value of energy transfer $\nu$ the integration domain in
momentum transfer $q$ is

\bel\begin{array}{l}\displaystyle
q\in
\Biggl( \sqrt{\nu^2-m^2+2E(E-\nu )-\rho},\\[4mm]
\displaystyle
\qquad {\rm min}\Bigl( \sqrt{\nu^2-m^2+2E(E-\nu)+\rho },\\[4mm]
\displaystyle \qquad
\sqrt{\nu^2-m_{\pi}^2+2M(\nu-m_{\pi})}\Bigr)\Biggr)
.\end{array}\ee where
$$\rho \equiv 2E\sqrt{(E-\nu )^2-m^2}.$$
It turns out that at above mentioned values of $\nu$ and $E$ the
allowed kinematical domain begins to decrease quickly with $\nu$
because the first argument in $min$ becomes smaller then the
second one.

vii) In the differential cross section of $\cos\theta_{\Delta}$
($\theta_{\Delta}$ is angle between $\Delta$ and incident neutrino
momenta) (Fig. 9, 10) one can notice maxima at values
$\cos\theta_{\Delta}\sim 0.875$ ($\theta_{\Delta}\sim 29^o$) for
$E_{\nu}=1 GeV$ (Fig. 9) and $\cos\theta_{\Delta}\sim 0.775$
($\theta_{\Delta}\sim 39^o$) for $E_{\nu}=2 GeV$ (Fig. 10). The
maxima are present in predictions of both models. For neutrino
energy $E_{\nu}=2 GeV$ (Fig. 12) there is is also a second maximum
in the forward direction. The shape of differential cross-sections
is similar to the one derived in \cite{singh}.

\bigskip

Our final conclusion is that pre-Marteau model for $\Delta^{++}$
excitation leads to close to standard behavior of $\pi^+$
production cross section and it is legitimate to use it in
sophisticated computations of nuclear effects and in MC codes .

\bigskip\bigskip

{\bf APPENDIX A}
\medskip

 A logic of the Marteau model can be
understood if one starts from the differential cross section for
quasi-elastic process $\nu_{\mu}\ n\rightarrow \mu^- p$ in the
Fermi gas model \cite{smithmoniz}:

\bel\begin{array}{l}\displaystyle d^2\sigma^{qel}_{FG}=\int d^3p
d\nu dq  \theta (k_F-|\vec p|)
\theta (|\vec p+\vec q|-k_F)\\[4mm]
\displaystyle
\qquad\qquad\qquad\times\delta (E+E_{\vec p}-E_{\vec
k'}-E_{\vec p'}) \\[4mm]
\displaystyle
\qquad\qquad\times{G^2\cos^2\theta_cN_{k_F}qM
M'\over 16E_{\vec p}E_{\vec p'}\pi E^2}
L_{\mu\nu}H^{\mu\nu}.
\end{array}\ee

$k^{\mu}$, $k'^{\mu}$, $p^{\mu}$, $p'^{\mu}$ denote $4$-momenta
of: neutrino, charged lepton, target and recoil nucleons. $M$ and
$M'$ are masses of target and ejected nucleons. In the case of
quasielastic process they are taken as equal.
$q^{\mu}=k^{\mu}-k'^{\mu}=(\nu , \vec q)$ is energy and momentum
transfer. $L_{\mu\nu}$ and $H^{\mu\nu}$ are leptonic and hadronic
tensors:

\bel\begin{array}{l}\displaystyle
L_{\mu\nu} = 8 \Bigl( k_{\mu} k'_{\nu} + k'_{\mu} k_{\nu} -
g_{\mu\nu}(k\cdot k')\\[4mm]
\displaystyle
\qquad\qquad\qquad + i\epsilon_{\mu\nu\alpha\beta}k'^{\alpha} k^{\beta} \Bigr)
\end{array}
\ee

\bel H^{\mu\nu} = {1\over 8MM'}Tr \Bigl(\Gamma^{\mu}(\hat p
+M)\tilde \Gamma^{\nu} (\hat p'+M')\Bigr),\ee

\bel
\tilde\Gamma^{\nu}=\gamma^0(\Gamma^{\mu})^{\dagger}\gamma^0,\ee

\bel\begin{array}{l}\displaystyle
\Gamma^{\mu}=F_1(Q^2)\gamma^{\mu}+iF_2(Q^2)\sigma^{\mu\nu}
{q_{\nu}\over
2M}
\\[4mm]
\displaystyle
\qquad +G_A(Q^2)\gamma^{\mu}\gamma_5+G_P(Q^2)\gamma_5{q^{\mu}\over 2M}.
\end{array}\ee

$F_1$, $F_2$, $G_A$ and $G_P$ are standard form-factors \cite{ls}.
$N_{k_F}={3A\over 8\pi k_F^3}={2V\over (2\pi)^3}$. We assume
nucleus of atomic number $A$ to contain equal numbers of protons
and neutrons.

In a good approximation
(since we will actually calculate cross section for a reaction on {\it free}
nucleons our derivation becomes exact;
we perform all the steps of computations starting from
the Fermi gas in order to check normalization factors) in ${1\over
E_{\vec p}E_{\vec p'}}H^{\mu\nu}$ we put $\vec p=0$ (thus $\vec
{p'}=\vec q$). The dependence on $\vec p$ factorizes and we
define
\bel\begin{array}{l}\displaystyle
\int d^3p \theta (k_F-|\vec p|) \theta (|\vec p+\vec
q|-k_F)\\[4mm]
\displaystyle
\qquad\qquad \times \delta (E+E_{\vec p}-E_{\vec k'}-E_{\vec p'}) \\[4mm]
\displaystyle \qquad\qquad = -{V\over 2N_{k_F}\pi}{\cal
I}m\Pi_{N-h}(\nu, \bar q).\end{array}\ee

The expression for the cross section takes form

\bel\begin{array}{l}\displaystyle d^2\sigma^{qel}_{FG}=-{\cal
I}m\Pi_{N-h}(\nu
,\vec q) \\[4mm]
\displaystyle \qquad\times {G^2\cos^2\theta_cM'qV\over
32\pi^2E^2E_{\vec q}}L_{\mu\nu}(H^{\mu\nu})_{\vec p=0}d\nu dq
\end{array}\ee

In the original Marteau approach non-relativistic nucleon's
kinematics is used and ${\cal I}m\Pi_{N-h}(\nu ,\vec q)$ is a
Lindhard function, the particle-hole polarization tensor, an
object which accounts for Fermi motion and Pauli blocking
\cite{fetter} (in \cite{marteau} higher order corrections in
$\frac{\vec p}{M}$ are considered).

In the limit $\vec p=0$ using non-relativistic decomposition of

\bel J^{\mu}=\overline {u}(\vec q)\Gamma^{\mu}u(\vec 0).\ee
one can identify in

\bel H^{\mu\nu}={1\over 2}\sum_{spins}J^{\mu}(J^{\nu})^*\ee
terms coming from different spin operators.
For example we calculated \cite{piony}:

\bel\begin{array}{l}\displaystyle
J^0=N_{p'}\phi_{s'}^{\dagger}\Bigl( {\bf
1}(F_1-F_2{q^2\over M'+E_{p'}})\\[4mm]
\displaystyle \qquad\qquad + \vec\sigma\cdot\vec q\frac{G_A-\nu
G_P}{M'+E_{p'}}\Bigl)\phi_s\end{array}\ee

where $N_{p'}=\sqrt{E_{p'}+M'\over 2M'}$ and $\phi$'s describe
non-relativistic spinors.

\medskip
Marteau $\Delta$ excitation model is defined by
\cite{marteau2}:
\smallskip

(i) substitution $M'=M_{\Delta}$;

(ii) multiplication of form-factors by the numerical factor
$4.78=(\frac{f_{\pi N\Delta}}{f_{\pi NN}})^2$;

(iii) elimination of "charge" terms (spin operator is necessary to
produce a particle of spin $3/2$) from $H^{\mu\nu}$ - we call the
new tensor $\tilde H^{\mu\nu}$ with the numerical factor $4.78$
included in its definition;

(iv) substitution of $\Pi_{N-h}$ by $\Pi_{\Delta - h}$ the
polarization tensor for $\Delta$ -hole excitation:

\bel\begin{array}{l}\displaystyle {\cal I}m\Pi_{\Delta -h}(\nu
,\vec q) = -{32\over 9}
{1\over (2\pi )^3}\int d^3p \theta (k_F-|\vec p|)\\[4mm]
\displaystyle
\qquad\qquad \times {M_{\Delta}^2\Gamma_{\Delta}\over
(W^2-M_{\Delta}^2)^2+M_{\Delta}^2\Gamma_{\Delta}^2}.\end{array}\ee

(the factor ${32\over 9}$ comes from summation over isospin and
spin degrees of freedom),

\bel W^2 = (E_{\vec p}+\nu )^2-(\vec p+\vec q)^2;\ee

(v) inclusion of RPA correlations and local density effects.

\smallskip
In our derivation we restricted ourselves to steps (i-iv) and we
obtained a model for $\Delta$ excitation on nuclei in the free
Fermi gas approximation:

\bel\begin{array}{l}\displaystyle d^2\sigma^{\Delta}_{FG}=-
{\displaystyle G^2\cos^2\theta_c
V M_{\Delta}q\over \displaystyle 32\pi^2E^2E_{\vec q}}\\[4mm]
\displaystyle \qquad\times {\cal I}m\Pi_{\Delta-h}(\nu , \vec
q)L_{\mu\nu}(\tilde H^{\mu\nu})_{\vec p=0}d\nu dq .\end{array}\ee

In the limit $k_F\rightarrow 0$ we obtained
pre-Marteau model (in this limit target nucleon
is at rest and our evaluation of
${H^{\mu\nu}\over E_{\vec p}E_{\vec p'}}$ becomes exact):

\bel\begin{array}{l}\displaystyle
{\cal I}m\Pi_{\Delta-h} \rightarrow -{8A\over
9V}\\[4mm]
\displaystyle \times {M_{\Delta}^2\Gamma_{\Delta}\over
\bigl((M+\nu )^2-\vec
q^2-M_{\Delta}^2\bigr)^2+M_{\Delta}^2\Gamma_{\Delta}^2}.\end{array}\ee

Finally (after $k_F$'s get properly cancelled)

\bel\label{DM}\begin{array}{l}\displaystyle
d^2\sigma^{\Delta}={G^2\cos^2\theta_cM_{\Delta}^3\Gamma_{\Delta}A
q\over 36\pi^2 E^2(M+\nu ) } \\[4mm]
\displaystyle \qquad\times {L_{\mu\nu}(\tilde H^{\mu\nu})_{\vec
p=0}d\nu dq\over \bigl((M+\nu )^2-\vec
q^2-M_{\Delta}^2\bigr)^2+M_{\Delta}^2\Gamma_{\Delta}^2}
\end{array}\ee $\Gamma_{\Delta}$ is defined as ($\Gamma_0=115MeV$)
\bel{}\Gamma_{\Delta}=\Gamma_0{q_{cm}(W)^3\over
q_{cm}(M_{\Delta})^3} {M_{\Delta}\over W},\ee where $q_{cm}(W)$ is
the pion momentum in $\Delta$ (of mass $W$) rest frame.

Pre-Marteau model provides a prediction for an overall $\Delta$
production i.e. for a sum over isospin degree of freedom. Without
nuclear effects relative probabilities to produce isospin states
is given by a ratio of Clebsh-Gordan coefficients. Thus for
neutrino induced reaction the probability to produce $\Delta^{++}$
is three times the probability to produce $\Delta^{+}$. In this
paper we present a comparison for $\Delta^{++}$ production. It is
because in the measurements of the invariant hadronic mass
distribution for the process $\nu_{\mu} n\rightarrow \mu^-p\pi^+$
there is a sharp resonance peak at $W\sim 1.2\ GeV$ while in the
channels $\nu_{\mu} p\rightarrow \mu^-n\pi^+$ and $\nu_{\mu}
p\rightarrow \mu^-p\pi^0$ peaks are smeared out \cite{exp}. It is
clear that correct description of the last two channels requires
an addition of non-resonant contribution or/and contributions from
other resonances while in the first one $\Delta^{++}$ production
cross section can be meaningfully compared with the data.
Prediction for $\Delta^{++}$ production per nucleon is thus
obtained by dividing (\ref{DM}) by ${A\over 2}$ and multiplying by
${3\over 4}$. The final formula for $\Delta^{++}$ excitation cross
section \emph{per proton} in the pre-Marteau model reads:

\bel\begin{array}{l}\displaystyle d^2\sigma_M^{\Delta^{++}}=
{G^2\cos^2\theta_cqM_{\Delta}^3\Gamma_{\Delta}\over
24\pi^2 E^2(M+\nu ) }\\[4mm]
\displaystyle \qquad\times {L_{\mu\nu}(\tilde H^{\mu\nu})_{\vec
p=0} d\nu dq\over \bigl((M+\nu )^2-\vec
q^2-M_{\Delta}^2\bigr)^2+M_{\Delta}^2\Gamma_{\Delta}^2}
\end{array}\ee

\bigskip
{\bf Acknowledgments}
\bigskip

The author (supported by KBN
grant 344/SPB/ICARUS/P-03/DZ211/2003-2005) thanks Krzysztof Graczyk for
useful conversations.

\end{document}